# Network Diffusion Model Reveals Recovery Multipliers and Heterogeneous Spatial Effects in Post-Disaster Community Recovery

Chia-Fu Liu[1*], Ali Mostafavi[2]

[1] Ph.D. Student, Zachry Department of Civil and Environmental Engineering, Texas A&M University, 199 Spence St., College Station, TX 77843-3136; e-mail: joeyliu0324@tamu.edu

[2] Associate Professor, Zachry Department of Civil and Environmental Engineering, Texas A&M University, 199 Spence St., College Station, TX 77843-3136; e-mail: amostafavi@civil.tamu.edu

**Abstract**

Community recovery from hazards and crises occurs through various diffusion processes within social and spatial networks of communities. Existing knowledge regarding the diffusion of recovery in community socio-spatial networks, however, is rather limited. To bridge this gap, in this study, we created a network diffusion model to characterize the unfolding of population activity recovery in spatial networks of communities. Using data related to population activity recovery durations calculated from location-based data in the context of 2017 Hurricane Harvey in the Houston area, we parameterized the threshold-based network diffusion model and evaluated the extent of homogeneity in spatial effects. Then we implemented the network diffusion model along with the genetic algorithm to simulate and identify recovery multipliers. The results show that the spatial effects of recovery are rather heterogeneous across spatial areas; some spatial areas demonstrate a greater spatial effect (spatial interdependence) in their recovery compared with others. Also, the results show that low-income areas demonstrate a greater spatial effect in their recovery. The greater spatial effects in recovery of low-income areas imply more reliance on resources and facilities of neighboring areas and also explain the existence of slow recovery hotspots in areas where socially vulnerable populations reside. Also, the results show that low-income and minority areas are community recovery multipliers; the faster the recovery of these recovery multipliers; the faster the recovery of the entire community. Hence, prioritizing these areas for recovery resource allocation could expedite the recovery of the overall community and promote recovery equality and equity.

**Significance Statement**

The study presented in this paper shows, for the first time, that post-disaster recovery follows a network diffusion process in which the extent of spatial effects (i.e., the extent of spatial interdependency) is highly heterogenous across spatial areas. The characterization of spatial spread of post-disaster community recovery based on a network diffusion model reveals significant insights about the spatial characteristics of recovery and provides important implications for recovery plans and implementations.

**1. Introduction**

The study examines the influence of socio-spatial network processes on post-disaster recovery. Recovery duration is one of the critical processes of community resilience to hazards and crises (Burton, 2015; Davidson & Cagnan, 2004; Miles & Chang, 2003). While existing studies have investigated determinants, such as the role of social capital (Cohen et al., 2013; Hikichi et al., 2020; Meyer, 2018) and socio-demographic attributes (Norris et al., 2008; Sherrieb et al., 2010) of community recovery in the aftermath of disasters, little is known about network effects in the progression of community recovery across time and space. In fact, community recovery unfolds upon the socio-spatial networks embedded in communities (Pezzica et al., 2020). Hence, understanding network processes that govern the diffusion of recovery in community socio-spatial networks is critical to leveraging network effects in expediting and improving post-disaster recovery. To bridge this gap, in this study, we use a network diffusion model to examine the extent of spatial contiguity effects in the spread of recovery among spatial areas of a community. We measure the empirical recovery duration of spatial areas based on the population activity recovery (Coleman et al., 2022), which uses the time for population visitations to points of interest to return to the pre-disaster patterns as a proxy for recovery. The network diffusion model is parameterized using empirical data from the 2017 Hurricane Harvey in the Houston metro area (Harris County, Texas). We use the model to evaluate the extent of spatial contiguity effects (i.e., spatial effects in the diffusion of recovery) and the extent to which spatial effects are homogeneous across the community. Along with the network diffusion model, we use the genetic algorithm (GA) to identify recovery multipliers: spatial areas whose recovery would expedite the recovery of the entire community. The specification of recovery multipliers would inform about ways recovery resources can be prioritized to spatial areas to expedite the recovery of the entire community.

The significance of socio-spatial networks embedded in communities has recently received growing attention, specifically in the context of disaster resilience. For example, Liu & Mostafavi (2022) examined hazard exposure heterophily in community spatial networks as a property that could influence community resilience. Lee et al. (2022) characterized important homophilic and heterophilic characteristics in communities' socio-spatial networks influencing population response to climate hazards. The findings of these studies demonstrate the significance of socio-spatial networks as structures upon which various community resilience processes unfold. Of significance for community resilience is the recovery process; however, the existing knowledge is missing information regarding the diffusion of recovery in community spatial networks. Specifying the existence and understanding the characteristics of network diffusion in post-disaster recovery can reveal insights regarding spatial effects and properties that could expedite and improve recovery. This knowledge gap motivated the research questions in this study.

In the selection of the network diffusion model, our work is inspired by the broader field of network contagion and diffusion modeling. Network diffusion has been studied in the context of the processes for spreading knowledge (Cowan & Jonard, 2004), virus transmission (Szor, 2005; Wang et al., 2013), ideas (Burt, 1987), and new behaviors (Valente, 1996) across social, natural, and physical networks. The study of diffusion phenomena has paved the way for a better understanding of the dynamics of complex networks in a variety of scientific fields. Various theoretical network diffusion models have been proposed. For instance, in the SIR model (Kermack et al., 1927), which is a classic model of epidemic spread, network diffusion is modeled based on various states a node can have throughout the duration of an epidemic: susceptible (S), infected (I), and recovered (R) states. Other epidemic spread models proposed by Kermack include the SI and SIS models (Kermack et al., 1927). The SI (susceptible–infected) model assumes that once a node becomes infected, it remains infected, while in the SIS (susceptible–infected–susceptible) model, an infected person has a probability of becoming susceptible again. Network diffusion models have been adopted in examining other network spread processes. An example is the spread of the digital virus, where the malicious software tries to transfer a copy of itself from one mobile device to another (Szor, 2005; Wang et al., 2013). Another example is the social contagion, which illustrates the spread of knowledge and innovation in a social network where people are deciding to adopt new concepts or innovations (Burt, 1987).

In addition to the SIR models and their variations, the other two main types of diffusion models include the threshold model (Granovetter, 1978) and the Independent Cascade model (Kempe et al., 2003). In this study, we use the threshold diffusion model in examining the spatial diffusion of recovery, which enables us to examine spatial contiguity effects. In the threshold model (first introduced by Granovetter (1978)), a node has two mutually exclusive states (binary states); it may elect to adopt a certain behavior, such as whether to take part in a riot. The node's state depends on the percentage of its neighbors (i.e., the threshold value) which have the same state. The model works as follows: each node starts with its own threshold and state (e.g., recovered and not recovered). During iteration $t$, each node is observed: if the percentage of its neighbors recovered at time $t - 1$ exceeds its threshold, it will also become recovered. Based on this model, we can estimate the threshold values for each spatial area based on the empirical recovery data and evaluate the extent of spatial contiguity effects across different spatial areas. The following section describes the details of the community recovery data and the threshold diffusion model of community recovery.

## 2. Results

### *2.1. Empirical threshold parameters*

In this section, we present the results of the first- and second-stage optimization. In the first stage, we implement the network diffusion model to specify the empirical threshold values. By applying the genetic algorithm (GA) on population activity recovery (measured by visits to points of interest) in Harris County after Hurricane Harvey, we obtain the empirical threshold values for each census block group (CBG). The results show that the parameterized model is able to capture the spatial effects among communities and then generate the recovery diffusion process through the underlying network structure. Figure 1 shows that the differences in the number of recovered CBGs per week between the empirical data and the simulated results are all below 100, except for the difference at week 14. The reason for the apparent difference (401) at week 14 is due to the research window in processing the population activity recovery duration. To eliminate seasonality effects (effects of holiday season) on population activities, any CBG recovery taking longer than 14 weeks, was designated as 14 weeks (3.5 months). Therefore, any CBG that has not recovered by the end of this window is considered to be recovered by the end of week 14 in the empirical data, which explains the spike in the difference at week 14. The results show that the model has a good ability to simulate and reconstruct the trajectory of the recovery process after the hurricane through the underlying network diffusion process.

### *2.2. Specifying recovery multipliers*

In the second-phase optimization, we use the parameterized diffusion model to specify the recovery multipliers in the simulated recovery process. A total of 1,609 CBGs were recovered at the end of week 14, leaving 401 CBGs in the *affected* states. We performed the GA to optimize the set of $\mathcal{M}^N$, which is defined as *recovered* at $\theta$ = 0, in four different cases: $N \in$ {20, 60, 101, 201}, which are {1%, 3%, 5%, 10%} of the CBGs throughout the county. The objective function is to maximize the number of recovered CBGs at the end of the diffusion process ($t$ = 14) by choosing the right set of $\mathcal{M}^N$. Figure 2 shows that the number of recovered CBGs would increase from 5.96% to 15.27% using the optimized recovery multipliers, which can significantly improve recovery in the entire region. As shown in Figure 3, we geographically identify the recovery multipliers in the four different cases. These CBGs are the critical recovery spreaders whose fast recovery would increase and expedite the recovery of the entire community.

### *2.3. Threshold value analysis*

By modeling post-disaster community recovery as a network diffusion process, we characterize the spatial effect of recovery among neighborhoods and spatial areas. In particular, the model shows that the spatial effects are heterogeneous, indicated by varying thresholds in our diffusion model. Essentially, the threshold represents the minimum percentage of recovered neighborhoods before the entire community is considered to be recovered. A higher threshold means more recovered neighbors are located around the CBG before it is recovered. This indicates that a clearer spatial effect was perceived around the CBG. The mean and variance of the threshold $\tau$ are 0.345 and 0.125, respectively. A greater spatial effect implies that there is more spatial interdependence for recovery among areas. Areas with low spatial effect could be resourceful and self-sufficient and thus able to recover with minimal reliance on the resources and facilities of neighboring areas. Such minimal reliance on resources and facilities of neighboring areas would create high spatial interdependence among areas for recovery that would lead to a higher spatial effect in recovery diffusion.

The variation in spatial effects did not show a significant statistical correlation with the extent of flooding in CBGs. We also examine the relationship between the spatial effect and the socio-demographic attributes of CBGs. The results show that CBGs in which stronger spatial effects are present, indicated by a higher threshold, have both lower per capita and lower household median income. As shown in Figure 4, the low-threshold group (first tertile) includes a higher percentage of both low per capita and household incomes. The middle-threshold group (second tertile) includes a higher percentage of lower per capita incomes than the high-threshold group (third tertile). The implication of this result is that CBGs with lower income (i.e., socially vulnerable groups) are areas where a stronger spatial effect needs to take place to effect their recovery. This finding implies that socially vulnerable areas are more dependent on resources and facilities of neighboring areas and hence have greater spatial interdependence for recovery (higher spatial effect in recovery diffusion). This finding could reveal one possible situation that delays the recovery of socially vulnerable areas: being surrounded by other socially vulnerable areas. The spatial effect for the diffusion of recovery that could create recovery isolates— hotspot clusters of CBGs from of socially vulnerable groups that struggle to recover due to absence of spatial effect from the neighboring areas.

*2.4. Recovery multiplier analysis*

In the next step, we examine the recovery multipliers in terms of their social-demographic attributes. Two notable findings were observed (Figure 5) from the analysis: (1) With a smaller number of multipliers selected, the identified multipliers are areas with lower per capita/household income; also in all four cases of multiplier sizes, multipliers have lower per capita/household income than the non-multiplier; and (2) In the case of smaller multiplier size, the multipliers have a larger minority percentage, and in all four cases, multipliers have a higher minority percentage than the non-multiplier. The two results imply that identifying the recovery multipliers and allocating resources to them not only improve the equality aspects of recovery for the vulnerable population but also play an important role in the diffusion of recovery in the entire community. These multipliers are the neighborhoods that need to be prioritized for resource allocation in the recovery period to improve both the equity and speed of recovery in the entire community. This result is significant since it shows improving equity in recovery resource allocation could benefit the recovery of the entire community.

**3. Discussion**

Post-disaster recovery is one of the critical processes of community resilience to hazards. Despite several years of research on the determinants of community recovery, little knowledge exists regarding the spatial network processes influencing community networks. Socio-spatial networks are structures upon which various processes related to community resilience unfold. In this study, we investigated community recovery as a network diffusion process in order to evaluate the extent of spatial contiguity effects in the spread of recovery across different areas of community and also identify ways to expedite the recovery of the entire community through activating network effects with the specification of recovery multipliers.

The findings of this study provide important scientific and practical contributions. This study is the first research to investigate network effects in post-disaster community recovery. While prior studies had identified the presence of disparities in recovery speed of areas, the majority of studies attribute the slow recovery to sociodemographic attributes. The findings of this study reveal the heterogeneity of spatial contiguity effects in community recovery and determine that low-income areas require a greater spatial effect for recovery compared with higher income areas. This finding implies that socially vulnerable areas are more dependent on resources and facilities of neighboring areas and hence have greater spatial interdependence for recovery, in other words, higher spatial effect in recovery diffusion. Also, prior studies have reported the presence of slow recovery hotspots in the communities; however, limited knowledge existed about mechanisms that contribute to clustering of slow recovery areas. Our findings resolve this unknown by showing that a CBG surrounded by other socially vulnerable areas and the absence of the required spatial effect for diffusion of recovery that could create recovery isolates (hotspot clusters of CBGs from socially vulnerable groups that struggle to recover due to the absence of spatial effect from the neighboring areas). In addition, the findings of this study introduce the novel concept of recovery multipliers and show that focusing on socially vulnerable areas as recovery multipliers not only enhances the equity of post-disaster recovery but also expedites the recovery for the entire community.

From a practical perspective, the model in this study could be used to proactively monitor community recovery and predict the diffusion of recovery. Also, the findings of this study have important implications for public officials, emergency managers, and decision makers involved in disaster recovery to better leverage the diffusion processes to accelerate and also promote equity in post-disaster community recovery. For example, identifying and prioritizing recovery multipliers for recovery resource allocation greatly enhance the speed of recovery in the entire community and also augment recovery equity.

As the first study focusing on network dynamics of community recovery, the findings of this study open avenues for subsequent areas of inquiry for future studies. For example, future studies can explore attributes of spatial areas (such as the built environment and hazard impacts) and their association with the spatial contiguity effect in community recovery diffusion. In addition, future studies can examine spatial effects and recovery diffusion across different hazard events and regions to evaluate the general characteristics of community recovery diffusion in spatial networks. Addressing the lines of inquiry focusing on recovery network dynamics would move us closer to better understanding and improving post-disaster recovery of communities through the lens of network processes.

## 4. Materials and Methods

### 4.1. Population Activity Recovery Data

In this study, we estimate the recovery of spatial areas based on population activity patterns. Prior studies (Coleman et al., 2022; Podesta et al., 2021) have shown that examining the fluctuations in population visitations to points of interest can provide important insights about the extent of impact and the duration of recovery of population activities. While population activity recovery may not capture all aspects of population post-disaster recovery, it can serve as a proxy indicating an important recovery milestone. Population activity fluctuations capture the combined effects of disruptions in households' lifestyles, infrastructure services, and businesses. Hence, the return of population activities to the pre-disaster steady state could be a reliable measure of aspects of community recovery. Also, through the use of location intelligence data, population activity recovery duration can be determined at fine spatial scales, such as at the census block group level. Such measurement of recovery duration at such a fine spatial scale is barely possible using standard methods of community recovery measurements based on surveys.

To specify the duration of population activity recovery, we followed the same data processing and method proposed in Coleman et al. (2022). We analyzed the de-identified and aggregated location data to characterize the pattern of population activity recovery in Harris County, Texas (USA) following Hurricane Harvey in 2017. Hurricane Harvey made landfall in August 2017, severely impacting the Texas coastal area. The extensive flooding rendered many essential and non-essential facilities inaccessible, significantly impacting post-disaster activities. Therefore, this scenario provides an appropriate context for studying the pattern of population recovery.

We first analyzed human mobility data obtained from Spectus, a commercial supplier of aggregated and de-identified mobility data, and calculated the daily number of visits by each census block group to various points of interest to specify fluctuations in population activity patterns. Second, POIs were categorized into essential (such as gasoline stations, grocery, and pharmacies) and non-essential POIs (such as banks, restaurants, and recreational centers). Third, the baseline was set for each group of POIs to measure population activity change due to Harvey. The baseline was obtained by averaging the number of visits from August 1 through August 21, 2017. Finally, the study calculated a 7-day moving average for each day after August 27, with 3 days before the target day and 3 days after the target day. The study defined a community as recovered when they observed post-disaster activities for 3 days at 90% of the baseline level. In this study, we use the population activity recovery time of CBGs in terms of visits to essential POIs to develop the network diffusion model. The analysis is based on aggregating the recovery duration of users in each CBG. CBGs under the minimum visit count of four are not included in the dataset, making 2,010 out of 2,144 CBGs in Harris County available for use in the implementation of the network diffusion model. In the next section, we will present the network diffusion model.

*4.2. Recovery network diffusion model configuration*

This section presents the steps in constructing and implementing the network diffusion model (Figure 6). To develop the network diffusion model for population activity recovery, we first construct the spatial network topology $G = (\mathcal{V}, \mathcal{E})$ for 2,010 CBGs in Harris County, Texas. To characterize the spatial structure, several spatial proximity criteria for spatial data include queen, rook, and bishop contiguity. Rook contiguity defines neighbors by the presence of a common edge and bishop contiguity defines neighbors by a common vertex between two spatial units, queen contiguity defines neighbors as spatial units that share a common edge or vertex. In this study, we used the queen neighborhood structure to construct the undirected graph by specifying the CBGs as vertices and the queen neighborhood as edges. As shown in Figure 7, there are a total of 2,010 vertices and 6,079 edges in Harris County's queen topological network, with an average node degree *k* of 6.049 and a graph density *d* of 0.00301.

In this study, we implemented NDlib (Network Diffusion Library) (Rossetti et al., 2018) to model the population activity recovery as a network diffusion process. NDlib is a Python package built upon the NetworkX Python library and is a framework designed to describe, simulate, and study diffusion processes in complex networks. We use the threshold model (Granovetter, 1978; Kempe et al., 2003) to simulate the trajectory of the community recovery process in the spatial network. In a threshold model, a node has two distinct and mutually exclusive states, which are *affected* and *recovered*. Each node's change in the state depends on the percentage of its neighbors that has the same state, which establishes a state change threshold. A set of initially recovered nodes and the threshold for each node are specified before the diffusion process begins. During a generic iteration, each node is observed: if the percentage of its recovered neighbor is greater than its threshold, it will also be recovered.

The model implemented for the recovery process works as follows. Each CBG $v_i$, it has a threshold $\tau_i \in [0, 1]$ throughout the diffusion process along with the status $S_t^{(i)} \in \{0: \textit{affected}, 1: \textit{recovered}\}$ at the end of week *t*. In particular, $t \in \{0, 1, \ldots, 14\}$ and *t* = 0 signifies the end of the hazard event (Harvey rainfall stopped), which represents the beginning of the recovery phase. In our model, all the CBGs are defined as affected at *t* = 0. Since the minimum recovery duration after Hurricane Harvey was 2.14 weeks (population activities were affected in all areas regardless of flood impacts), we set the 750 CBGs whose recovery duration is less than 3 weeks to have threshold $\tau_i$ equal to zero, making them the first group of CBGs to recover at the end of week 3. The goal of the recovery diffusion model is to simulate the observed recovery duration of each CBG through the underlying network structure and estimate the threshold values for each CBG. The greater the threshold for a CBG, the higher the spatial effects in recovery of the CBG. The spatial effect captures the extent of spatial interdependence among the areas during recovery which is the extent to which an area relies on resources and facilities of neighboring areas for recovery. Fig 8 shows the progression of the recovery diffusion process.

The objective function of the optimized diffusion model is

$$\min_{\tau} L(S_t^{(i)}, \hat{S}_t^{(i)}) \tag{1}$$

where $S_t^{(i)}$ is CBG $v_i$'s state at time *t* observed from the empirical data while $\hat{S}_t^{(i)}$ is CBG $v_i$'s state at time *t* generated from the diffusion model. $L(\cdot)$ is defined as a 0-1 loss function

$$L(S_t^{(i)}, \hat{S}_t^{(i)}) = \sum_{t=1}^{14} \sum_{i=1}^{2010} I(S_t^{(i)} \neq \hat{S}_t^{(i)}) \tag{2}$$

where $I(\cdot)$ is the indicator function defined as

$$I(\cdot) := \begin{cases} 1 & \text{if } S_t^{(i)} \neq \hat{S}_t^{(i)} \\ 0 & \text{if } S_t^{(i)} = \hat{S}_t^{(i)} \end{cases} \tag{3}$$

Mathematically, our goal is to obtain a set of threshold values $\tau$ which will generate minimal loss (based on observed recovery durations) throughout the entire diffusion process. The notations in the diffusion model and their descriptions are summarized in Table 1.

*4.3. First step optimization— obtain parameterized diffusion model*

In the first stage optimization, we have 2,010 variables $\{\tau_1, \tau_2, \ldots, \tau_{2010}\}$ and each $\tau_i$ could take any real number between 0 and 1. Obviously, solving this problem would be prohibitively expensive. To solve this problem efficiently, we applied a genetic algorithm (GA) to solve the optimization problem. GA is an optimization algorithm used to solve challenging real-world problems encountered in various fields (Ghaheri et al., 2015; He et al., 2012; Peerlinck et al., 2019; Sari & Tuna, 2018).

In this study, we optimized the recovery diffusion model by classic GA described in Table 2. A series of thresholds $\{\tau_1, \tau_2, \ldots, \tau_{2010}\}$ form the chromosome representation. When GA optimizes the parameterized model, a population size $\eta$ of chromosomes will be randomly generated at iteration $\theta$ = 0. The population size of $\eta$ is maintained across the generic iteration. The fitness function is equal to the loss function $L(S, \hat{S})$ and is calculated for each chromosome at the end of the diffusion process (*t* = 14).

GA as a population-based metaheuristic algorithm has a proven ability to avoid becoming stuck in the local minimum. The larger the population size $\eta$, the greater the chance that a global optimal solution can be found. However, the programming runtime also increases significantly with increasing $\eta$. To find a promising population size $\eta$ to get a high-quality solution with tolerable runtime, we compared three different population sizes $\eta$ equal to 10, 15, and 20 under the maximum number of 2,500 iterations. Due to the stochastic nature of the formation of the initial population, we also performed randomized diffusion processes 1,000 times to get the loss $L(S, \hat{S})$ equal to 6569.835 to set as the baseline. To compare their performance, we define the algorithm performance $P^{(\eta)}$ as

$$P^{(\eta)} = \frac{\Delta L^{(\eta)}}{R^{(\eta)}} \quad (4)$$

where $\Delta L^{(\eta)}$ represents the loss descent rate per generation and $R^{(\eta)}$ represents the runtime (in seconds) required for each generation. As shown in Figure 9, at $\eta$ = 20, the algorithm has the lowest loss with, but also needs the longest runtime among three cases and thus generates the lowest $P^{(\eta)}$. On the other hand, the algorithm achieves the best performance index $P^{(\eta)}$ with a population size $\eta$ of 10. Therefore, in this study, we will set the population size $\eta$ = 10 and *MAX* = 10,000 to generate the optimal thresholds $\tau^* = (\tau_1^*, \ldots, \tau_{2010}^*)$ for 2,010 CBGs in Harris County, Texas. The optimization algorithm yields the final loss $L(\cdot)$ equals to 2,752.

*4.4. Second step optimization— identify recovery multiplier*

After obtaining the parameterized threshold model through the genetic algorithm, we want to identify the critical spatial areas (CBGs), which we call the recovery multipliers, whose recovery would expedite the recovery of the entire community. The idea here is that by selecting a set of CBGs and defining them as *recovered* from the beginning of the diffusion process, the model can finish the recovery process in the shortest period of time. In essence, by strengthening a specific subset of the CBGs in the spatial network topology, we can expedite the recovery of the entire region. In the diffusion model, we set recovery multipliers of size *N* ($\mathcal{M}^N$) as *recovered*, and simulate the network diffusion process by the parameterized model we learned from the first stage optimization. We selected four different sizes of recovery multiplier: *N* equals {20, 60, 101, 201} representing {1%, 3%, 5%, 10%} of the total CBGs. We set the maximum number of generic iterations *MAX* to 2,000 in the second stage of optimization.

**Data availability:**

The data used in this study are not publicly available under the legal restrictions of the data provider. Interested readers can request it from Spectus directly.

**Code availability:**

The code that supports the findings of this study is available from the corresponding author upon request.

**Acknowledgments**

This material is based in part upon work supported by the National Science Foundation under Grant CMMI-1846069 (CAREER). The authors also would like to acknowledge the data support from Spectus. Any opinions, findings, conclusions, or recommendations expressed in this material are those of the authors and do not necessarily reflect the views of the National Science Foundation and Spectus.

**Author Contributions:** C.F.L. and A.M. conceived the idea. C.F.L collected the data and carried out the analyses. C.F.L. and A.M. wrote the manuscript.

**Competing Interest Statement:** The authors declare no competing interests.

**Figures and Tables**

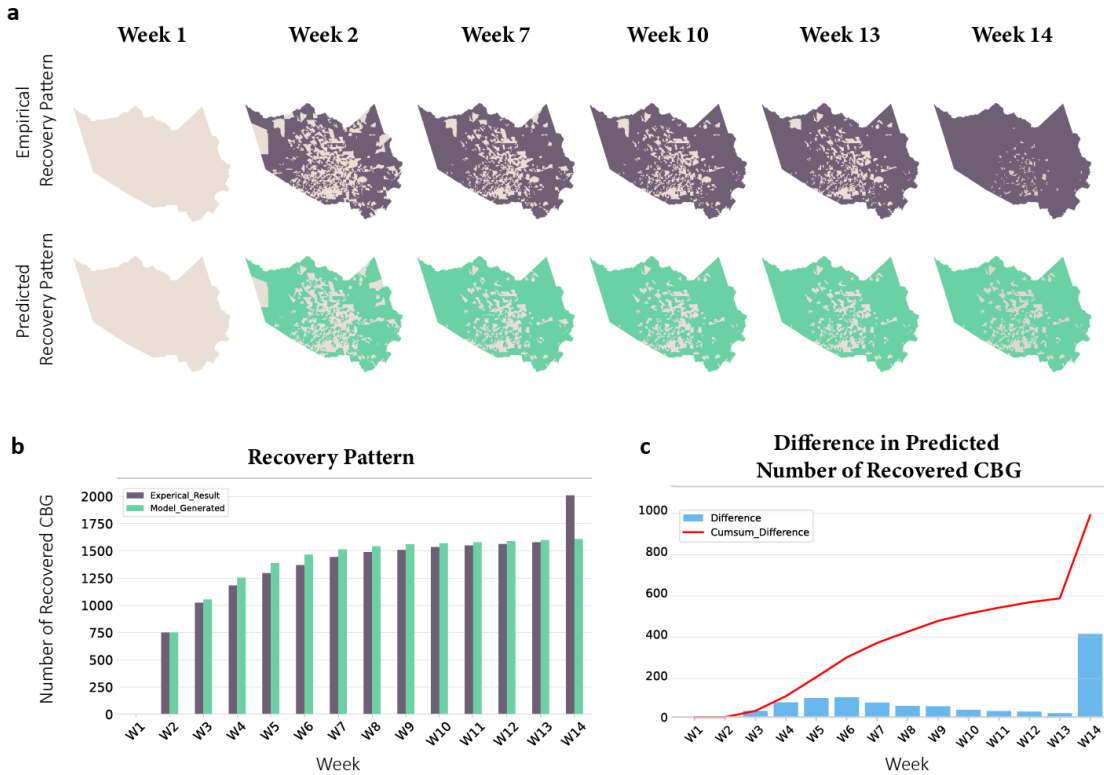

**Figure 1.** (*a*) The recovery pattern simulated by the network diffusion model on the activity data. (*b*) The number of recovered CBG in each week obtained from the empirical data versus diffusion model. (*c*) The difference and the accumulated difference in recovered CBG between the empirical data and simulated data.

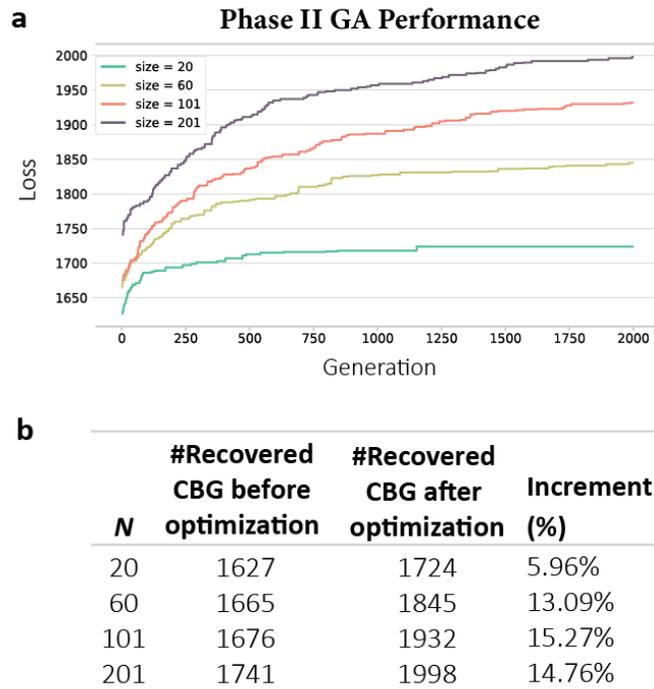

| N | #Recovered CBG before optimization | #Recovered CBG after optimization | Increment (%) |
|---|---|---|---|
| 20 | 1627 | 1724 | 5.96% |
| 60 | 1665 | 1845 | 13.09% |
| 101 | 1676 | 1932 | 15.27% |
| 201 | 1741 | 1998 | 14.76% |

**Figure 2.** (*a*) The algorithm performance in terms of the number of recovered CBG at the end of week 14. (*b*) The number of recovered CBGs before and after the optimization and the increment rate ($\frac{\#recovered\ with\ optimization - \#recovered\ without\ optimization}{\#recovered\ without\ optimization} \times 100\%$).

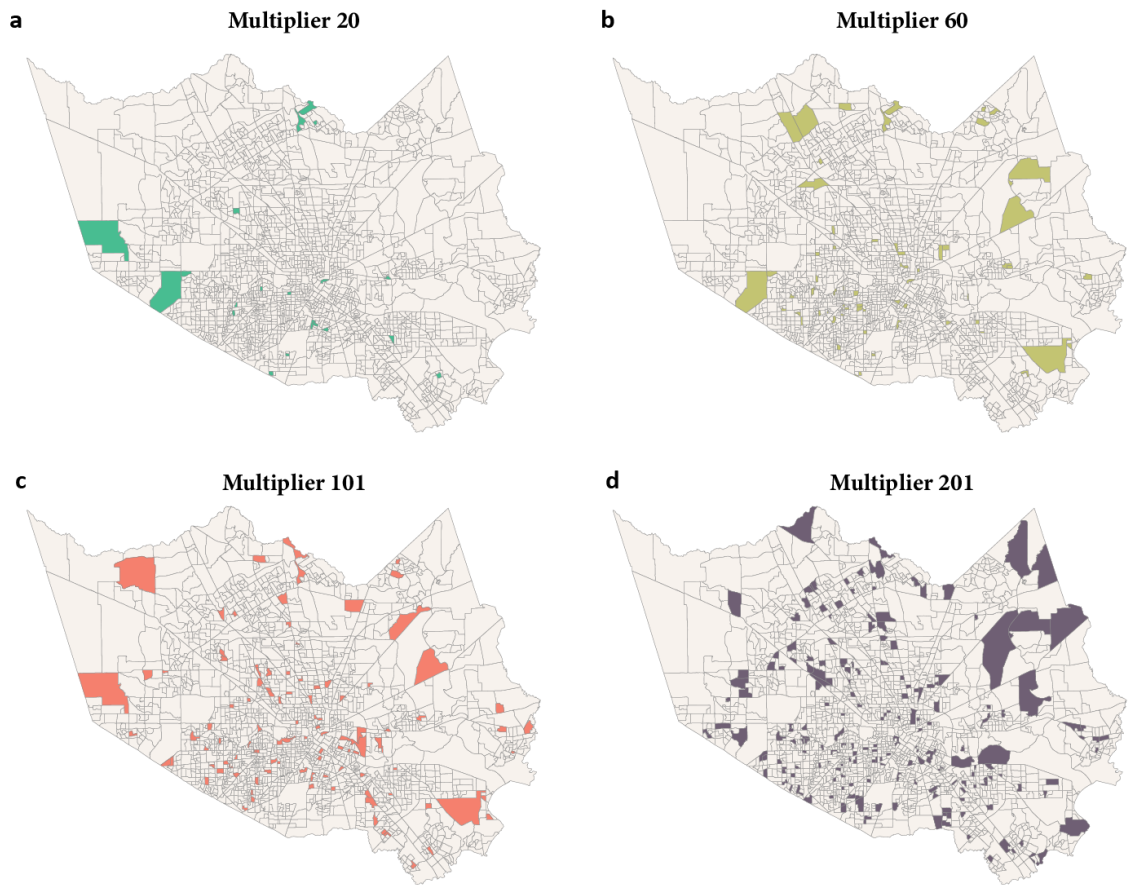

**Figure 3.** (*a*) – (*d*) corresponds to the geography of the optimized set $\mathcal{M}^N$ with *N* equal to 20, 60, 101, and 201.

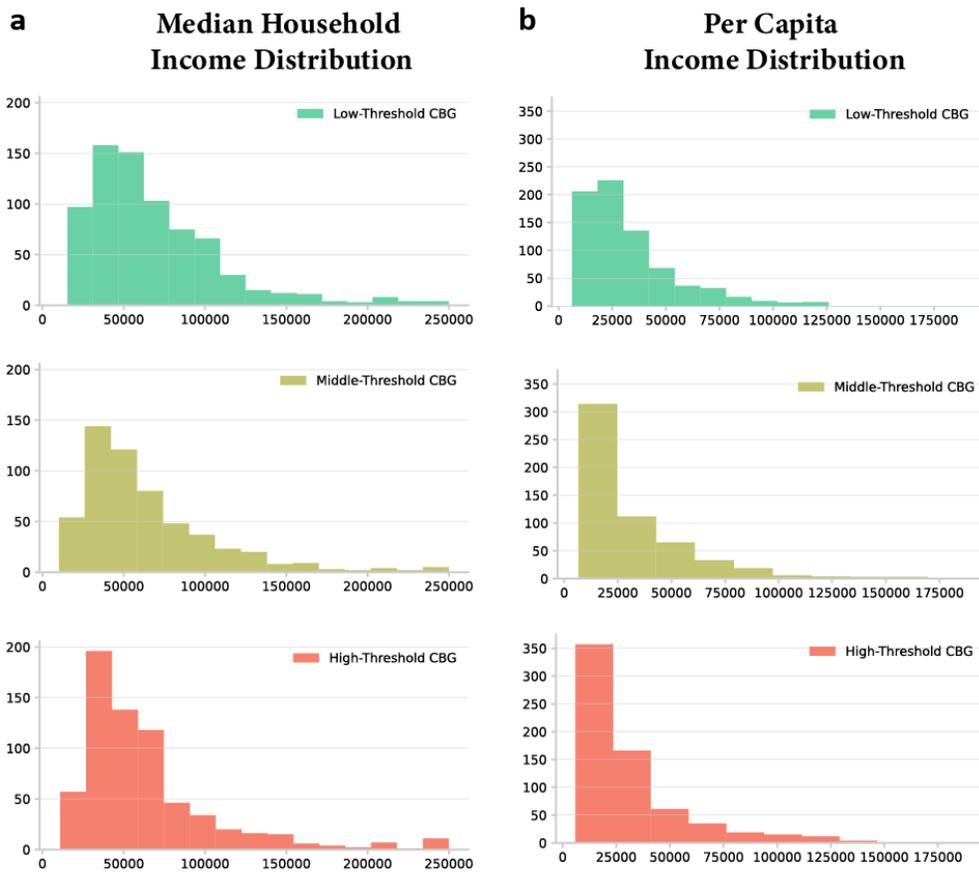

**Figure 4.** (*a*) The median household income distribution among low-threshold (first tertile), middle-threshold (second tertile), and high-threshold (third tertile) CBGs. (*b*) The per capita income distribution among low-threshold (first tertile), middle-threshold (second tertile), and high-threshold (third tertile) CBG.

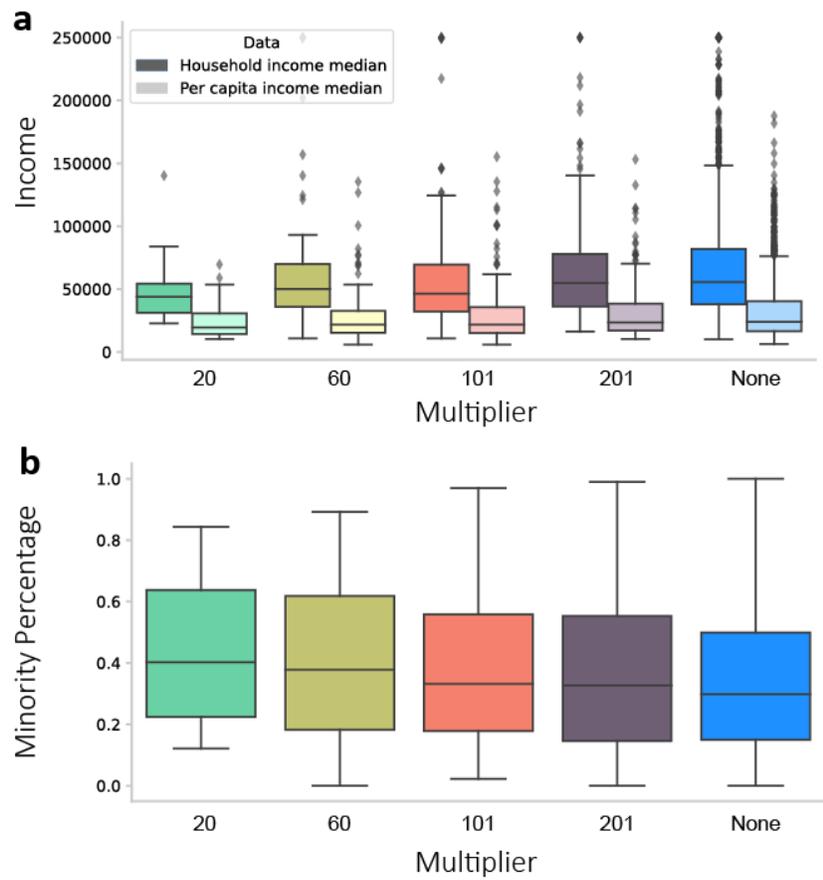

**Figure 5.** (*a*) The household and per capita income median distribution of the recovery multiplier with different sizes. (*b*) The minority percentage distribution of the recovery multiplier with different sizes.

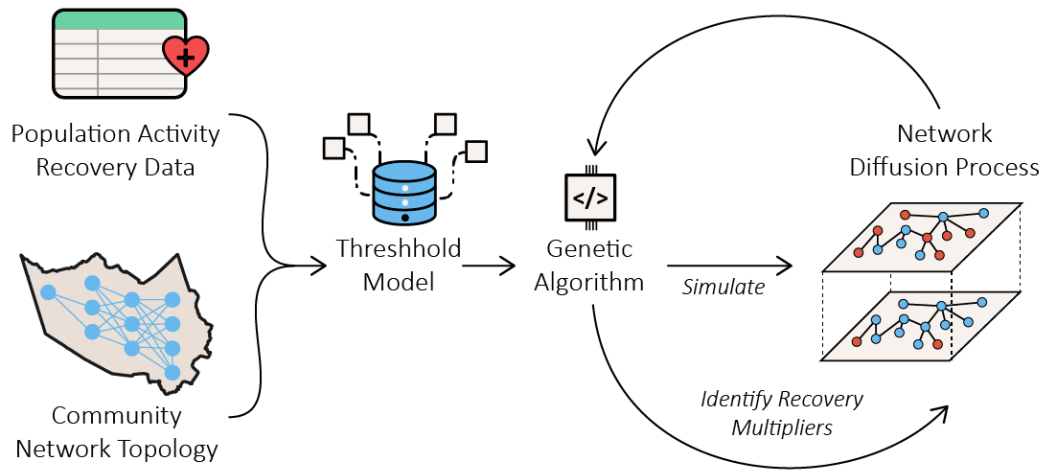

**Figure 6.** Overview of the steps for modeling post-disaster recovery as a network diffusion process.

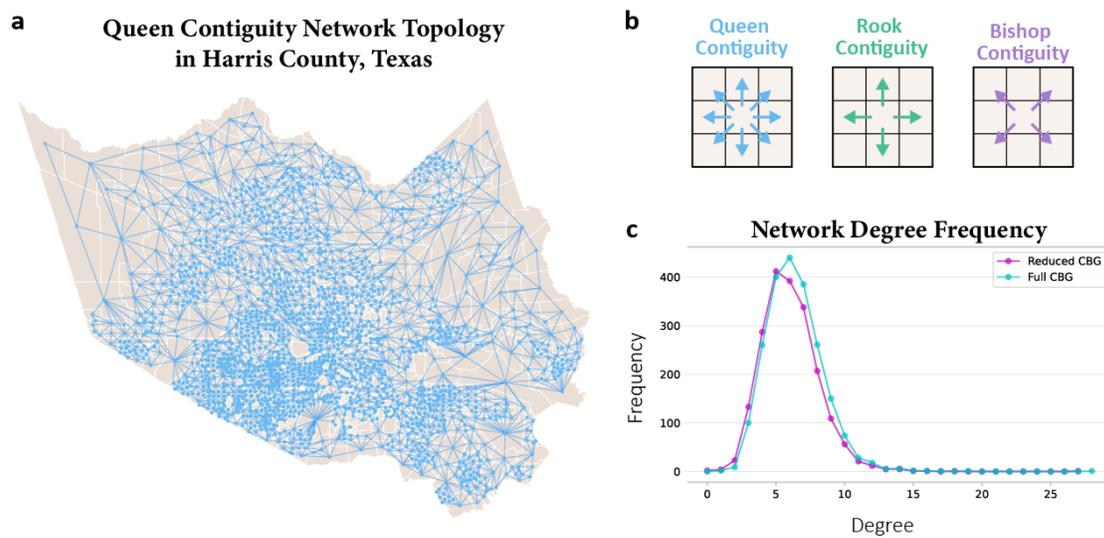

**Figure 7.** (*a*) Queen contiguity spatial network topology for the 2,010 CBGs in Harris County, Texas. (*b*) The illustration of queen, rook, and bishop contiguity. (*c*) The node degree distribution for full CBG (2,144) and reduced CBG (2,010) in Harris County, Texas.

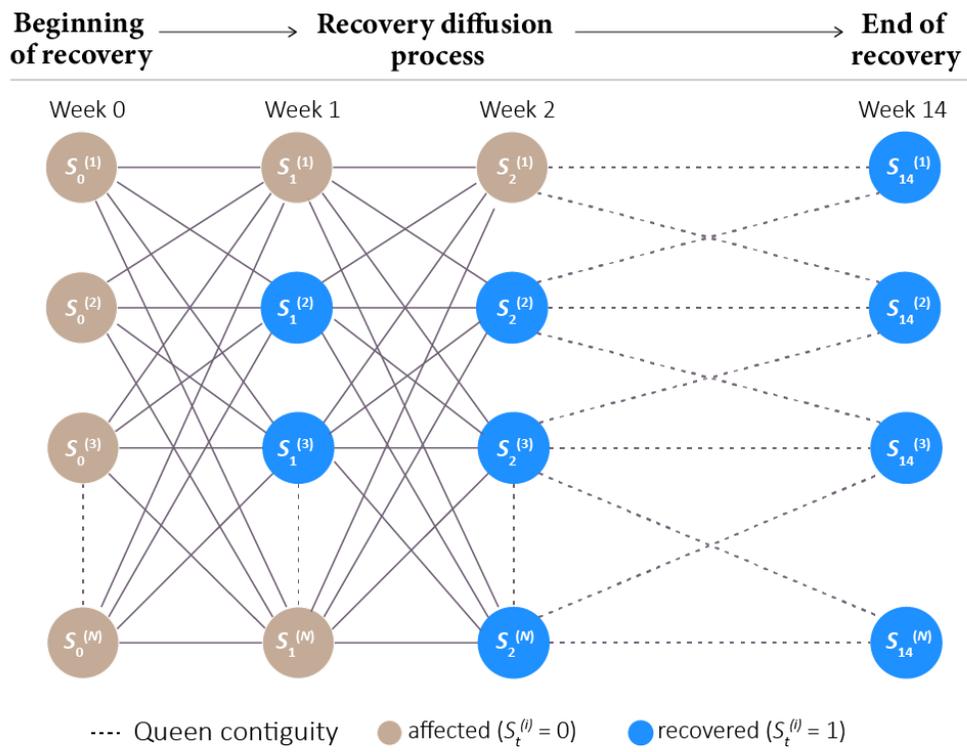

**Figure 8.** The recovery diffusion process implemented by the threshold model. A CBG $v_i$ would become recovered at week $t$ + 1 if $\frac{\sum_{j \in \mathcal{N}(v_i)} S_t^{(j)}}{\#\mathcal{N}(v_i)} \geq \tau_i$, where $\mathcal{N}(v_i)$ is set of the neighbor of $v_i$.

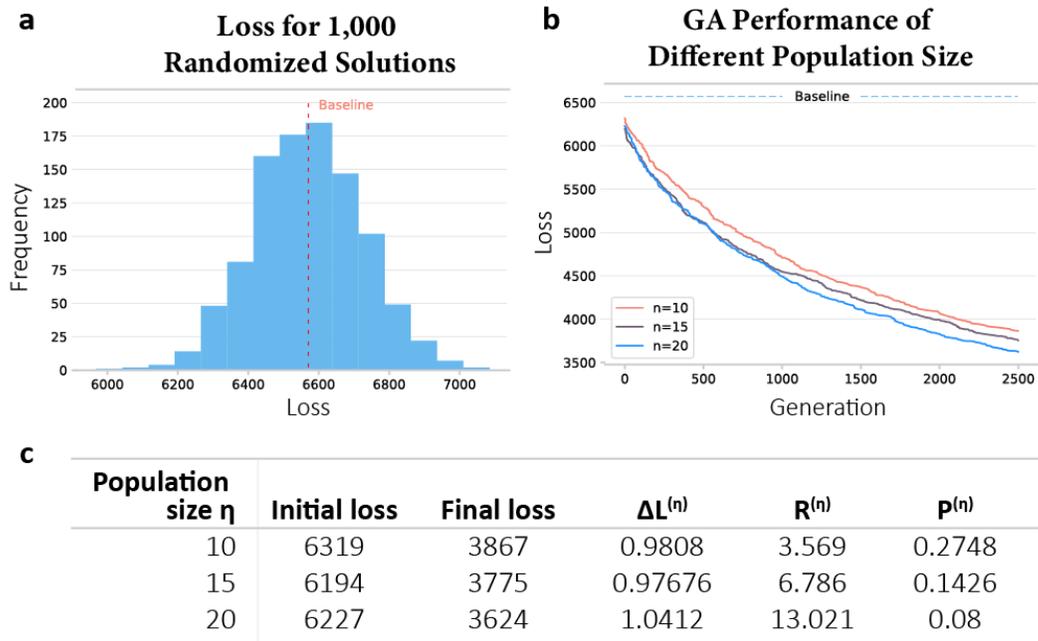

| Population size η | Initial loss | Final loss | ΔL$^{(n)}$ | R$^{(n)}$ | P$^{(n)}$ |
|---|---|---|---|---|---|
| 10 | 6319 | 3867 | 0.9808 | 3.569 | 0.2748 |
| 15 | 6194 | 3775 | 0.97676 | 6.786 | 0.1426 |
| 20 | 6227 | 3624 | 1.0412 | 13.021 | 0.08 |

**Figure 9.** (*a*) The loss of 1,000 randomized recovery diffusion processes and the mean as the baseline. (*b*) The loss of the algorithm with population size $\eta$ equal to 10, 15, and 20. (*c*) The algorithm performances with population size $\eta$ equal to 10, 15, and 20.

**Table 1.** List of major symbols and descriptions in the network diffusion model

| Sym. | Domain | Description |
|---|---|---|
| $\mathcal{V}$ | $\mathbb{N}^n$ | A set of vertices with size *n* |
| $\mathcal{E}$ | $\mathbb{N}^m$ | A set of edges with size *m* |
| $G = (\mathcal{V}, \mathcal{E})$ | $\mathbb{N}^{n \times m}$ | A network defined as an undirected graph with $\mathcal{V}$ and $\mathcal{E}$ |
| $\mathcal{N}(v)$ | $\mathbb{N}$ | The set of neighbor of vertex $v$ |
| $k$ | $\mathbb{R}$ | Average node degree in a graph |
| $d$ | $\mathbb{R}$ | Density for an undirected graph |
| $\tau_i$ | [0, 1] | Threshold value of $v_i$ in diffusion model |
| $S_t^{(i)}$ | {0, 1} | State of $v_i$ at the end of week *t* observed from the dataset |
| $\hat{S}_t^{(i)}$ | {0, 1} | State of $v_i$ at the end of week *t* generated from the model |

**Table 2.** Classical Genetic Algorithm (GA)

*Input:*
   Population Size, $\eta$
   Maximum number of iterations, *MAX*
*Output:*
   The global best solution, $\tau^*$

*begin*
   Generate initial population of *n* chromosome $\tau_i$ (*i* = 1, 2, …, *n*)
   Set iteration counter $\theta = 0$
   Compute the fitness value for each chromosome
   *while* ($\theta < MAX$)
      Select a pair of chromosomes from initial population based on fitness
      Apply crossover operation on selected pair with crossover probability
      Apply mutation on the offspring with mutation probability
      Replace old population with newly generated population
      Increment the current iteration $\theta$ by 1.
   *end while*
   return the best solution, $\tau^*$
*end*